\documentclass[12pt,english]{article}
\usepackage{graphicx}

\begin{document}

\title{The Two-State Vector Formalism}


\maketitle


\vskip 20 pt

The two-state vector formalism (TSVF) \cite{review} is a
time-symmetric description of the standard quantum mechanics
originated in Aharonov, Bergmann and Lebowitz \cite{ABL}. The TSVF
describes a quantum system at a particular time by two quantum
states: the usual one, evolving forward in time,  defined by the
results of a complete measurement at the earlier time, and by the
quantum state evolving backward in time, defined by the results of a
complete measurement at a later time.

 According to the standard quantum
formalism, an ideal (von Neumann) measurement  at time $t$ of a
nondegenerate variable $A$ tests for existence at this time of the
forward evolving state $|A=a\rangle$ (it yields the outcome $A=a$
with certainty if this was the state) and creates the state evolving
towards the future:
\begin{equation}\label{fest}
   |\Psi (t') \rangle=
   e^{-{i\over\hbar}\int_t^{t'}Hdt}|A=a\rangle,~~~t'>t.
\end{equation}
(In general, the Hamiltonians $H(t)$ at different times do not
commute and a time ordering has to be performed.)

 In the TSVF this
ideal measurement also tests for backward evolving state arriving
from the future $~\langle A=a |$ and creates the state evolving
towards the past:
\begin{equation}\label{best}
 \langle \Phi (t'')|=~
   \langle A=a| e^{{i\over\hbar}\int_{t}^{t''} Hdt},~~~t''<t.
\end{equation}

Apart from some differences (discussed below) following from the
asymmetry of the memory arrow of time, one can perform similar
manipulations of the forward and backward evolving states. In
particular,  neither can be cloned and both can be teleported.

Given complete measurements, $|A=a\rangle$ at $t_1$ and
$|B=b\rangle$ at $t_2$,    the complete description of a quantum
system at time $t$,  $t_1<t<t_2$, is
 the {\em two-state vector} \cite{AV90}:
\begin{equation}\label{tsv}
   \langle \Phi |~~|\Psi\rangle ,
\end{equation}
where the states $\langle \Phi |$ and $|\Psi\rangle$ are obtained
using (\ref{fest},\ref{best}).

 The two-state vector provides the maximal information
regarding the way the quantum system can affect at time $t$ any
other system. In particular, the two-state vector describes the
influence on a measuring device coupled with the system at time $t$.
An ideal measurement of a variable $O$ yields an eigenvalue $o_n$
with probability given by the Aharonov, Bergman, Lebowitz (ABL)
rule:
\begin{equation}
  \label{ABL}
 {\rm Prob}(o_n) = {{|\langle \Phi | {\bf P}_{O=o_n} | \Psi \rangle |^2}
\over{\sum_j|\langle \Phi | {\bf P}_{O=o_j} | \Psi \rangle |^2}} .
\end{equation}
This is, essentially, a conditional probability. We consider an
ensemble of pre- and post-selected quantum systems with the desired
outcomes of the measurements at $t_1$ and $t_2$. Only those systems
(and all of them) are taken into account.
 Intermediate measurement
(or the absence of it) might change the probabilities of the
outcomes of the post-selection measurement at time $t_2$, but this
is irrelevant: it only changes the size of the pre- and
post-selected ensemble given the size of the pre-elected ensemble at
$t_1$.

Note that the ABL rule simplifies the calculation of probabilities
of the outcome of intermediate measurements. In the standard
approach we need to calculate the time evolutions between time $t$
and $t_2$ of all states corresponding to all possible outcomes of
the intermediate measurement, while in the TSVF we have to calculate
evolution of only one (backward evolving) state.

The pre- and post-selected quantum system (described by the
two-state vector) has very different features relative to the system
described by a single, forward evolving quantum state. The
Heisenberg Principle does not hold: noncommuting observables might
be simultaneously well defined, i.e. each observable might have a
dispersion-free value provided that it was the only one measured at
time $t$. As an example, consider a spin-$1\over 2$ particle in a
field free region. Assume that  $\sigma _z$ was measured at $t_1$,
$\sigma _x$ at $t_2$ and both were found to be 1. When at time $t$,
an outcome of a measurement of a variable (if measured) is known
with certainty, it is named {\em an element of reality} \cite{ER}.
Thus, in the above example, both $\sigma _z=1$ and $\sigma _x=1$ are
such elements of reality.

For pre- and post-selected systems there might be apparently
contradicting elements of reality. Consider now a spin-$1\over 2$
particle which can be located in two boxes, $A$ and $B$, which is
described by the two-state vector:
\begin{equation}
 \langle \Phi |~~|\Psi\rangle = \frac{1}{3}
\left(\left\langle A,\uparrow_{z}\right|+\left\langle
A,\downarrow_{z}\right|-\left\langle
B,\uparrow_{z}\right|\right)~~~~\left(\left|A,\uparrow_{z}\right\rangle
 +\left|A,\downarrow_{z}\right\rangle
+\left|B,\uparrow_{z}\right\rangle \right)
,\label{eq:spin_post}\end{equation}
 (where $\left|A,\uparrow_{z}\right\rangle $ represents the particle
in box $A$ with spin $\uparrow_{z}$). Then, there are two elements
of reality: ``the particle in box $A$ with spin up'' and ``the
particle in box $A$ with spin down''. Indeed, the measurement of the
projection $\textbf{P}_{A\uparrow}$ has the outcome
$\textbf{P}_{A\uparrow}=1$ with certainty, and the outcome of the
other projection (if measured instead) is also certain:
$\textbf{P}_{A\downarrow}=1$. This can be readily verified using the
ABL rule or the standard formalism.

 Obviously,  the measurement of the
product of the projections is certain too:
$\textbf{P}_{A\uparrow}~\textbf{P}_{A\downarrow}=0$, so this example
shows also
 {\em the failure of the product rule}: at time $t$ we know with
certainty that if $A$ is measured, the outcome is $a$, and if $B$ is
measured instead, the outcome is $b$, but nevertheless, the
measurement of $AB$ is not $ab$. (The product rule does hold for the
standard, pre-selected quantum systems.)

This example is mathematically equivalent to the three-box paradox
\cite{AV91} in which a single pre- and post-selected particle can be
found with certainty both in box $A$ if searched there and in box
$B$ if searched there instead. These bizarre properties of elements
of reality generated much controversy about the {\em counterfactual}
usage of the ABL rule (see entry Counterfactuals in Quantum
Mechanics). It should be stressed that ``elements of reality''
should not be understood in the ontological sense, but only in the
operational sense, given by their definition.

The most important outcome of the TSVF is the discovery of {\em weak
values} of physical variables \cite{s-100}. When at time $t$,
another system couples weakly to a variable $O$ of a pre- and
post-selected system  $\langle \Phi |~~|\Psi\rangle$,  the effective
coupling is not to one of the eigenvalues, but to the weak value:
\begin{equation}
O_w \equiv { \langle{\Phi} \vert O \vert\Psi\rangle \over
\langle{\Phi}\vert{\Psi}\rangle } . \label{wv}
\end{equation}
The weak value might be far away from the range of the eigenvalues,
and this can lead to numerous surprising effects  described in the
entry ``Weak Value and Weak Measurement''.

There is an important connection between weak and strong
measurements. If the outcome of a strong measurement $O=o_i$  is
known with certainty, the weak measurement has to yield the same
value, $O_w=o_i$. The inverse is true for dichotomic variables: if
the weak value is equal to one of the two eigenvalues, a strong
measurement should give this outcome with certainty.

 In both strong
and weak measurements, the outcome manifests via the shift of the
pointer variable. For strong measurements it might be random, but
for weak measurements it is always certain (and equals to the weak
value). Sometimes it is called ``weak-measurement elements of
reality'' \cite{WMER}.

A generalization of the concept of the two-state vector (with
natural generalizations of the ABL rule and weak value) is a
``superposition'' of two-state vectors which is called a {\em
generalized two-state vector} \cite{AV91}:
\begin{equation}
  \label{g2sv}
 \sum_i  \alpha_i  \langle\Phi_i |~ |\Psi_i \rangle .
\end{equation}
A quantum system described by a generalized two-state vector
requires pre- and post-selection of the system together with an
ancilla which is not measured between the pre- and post-selection.

Systems described by generalized two-states vectors might have more
unusual properties. The Heisenberg uncertainty principle breaks down
in even more dramatic way: we can have a set of many noncommuting
observables having dispersion-free values and not just the trivial
case of two, one observable defined by pre-selection and another by
 post-selection. An extensively analyzed example of this kind is
``the mean king problem'' \cite{xyz,mking} in which we have to know
all observables of the set of the noncommuting observables for all
possible outcomes of the post-selection measurement.

Another natural multiple-time nonlocal generalization is to consider
$2N$-state vector (or generalized $2N$-state vector) which provides
a complete description of how a (composite) system can affect other
systems coupled to it in $N$ space-time points. Preparing and
testing such $2N$-state vectors require multiple-time and nonlocal
measurements. (Note that causality puts some constrains on such
measurements \cite{VN}.) An incomplete description in which we
associate only one (forward or backward) evolving state with some
space-type points is also of interest. For example,  two
spin-$1\over 2$ particles in an entangled ``state'' which evolves
forward in time for one particle and backward for the other
particle, can be completely correlated:
\begin{equation}
\frac{1}{\sqrt 2}\left(\left|\uparrow\right\rangle _{A}\left\langle
\uparrow\right|_{B}+\left|\downarrow\right\rangle _{A}\left\langle
\downarrow\right|_{B}\right) .\label{compcor}
\end{equation}
Here, the measurements of the spin in components in any direction
yield the same result for both particles. There is no pre-selected
quantum system with such property.

The TSVF is a time symmetric approach. However, there are some
differences between forward and backward evolving quantum states: we
can always create a particular forward evolving quantum state, say
$|A=a\rangle$. We measure $A$, and if the outcome is a different
eigenvalue than $a$, we perform an appropriate transformation to the
desired state. We cannot, however, create with certainty a
particular backward evolving quantum state, since the correction has
to be performed before we know the outcome of the measurement. The
difference follows from the time asymmetry of the memory arrow of
time. This asymmetry is not manifest in the ABL rule and the weak
value, because the outcome of measurement is the {\em shift} of the
pointer during the measurement interaction and this is invariant
under changing the direction of time evolution. The shift is between
zero and the outcome of the measurement and this is where the memory
arrow of time introduces the asymmetry. The state ``zero'' is always
in the earlier time: we do not ``remember'' the future and thus we
cannot fix the final state of the measuring device to be zero.

The TSVF is equivalent to the standard quantum mechanics, but it is
more convenient for analyzing the pre-and post-selected systems and
allowed to see numerous surprising quantum effects. The TSVF is
compatible with almost all interpretations of quantum mechanics but
it fits particularly well the many-worlds interpretation. The
concepts of ``elements of reality'' and ``weak-measurement elements
of reality'' obtain a clear meaning in worlds with particular
post-selection, while they have no ontological meaning in the scope
of physical universe which incorporates all the worlds.

\vskip .5cm Lev Vaidman\hfill\break School of Physics and Astronomy
\hfill\break Raymond and Beverly Sackler Faculty of Exact
Sciences\hfill\break Tel-Aviv University, Tel-Aviv 69978, Israel

\end{document}